\documentclass[twocolumn,prd,nofootinbib,superscriptaddress]{revtex4-2}
\usepackage{graphicx}
\graphicspath{ {./figures/} }

\usepackage{color}
\usepackage{bm}

\newcommand\unit[1]{{\rm #1}}
\newcommand\prx{PRX}
\newcommand\apjl{ApJL}
\newcommand{\RIFT}{RIFT}
\newcommand{\RM}{RUNMON-RIFT}
\begin{document}

\title{\RM{}: Adaptive Configuration and Healing for Large-Scale Parameter Inference}
\author{R. Udall}
\affiliation{LIGO Laboratory, Caltech}
\affiliation{Georgia Institute of Technology}
\author{J. Brandt}
\affiliation{Georgia Institute of Technology}
\author{G. Manchanda}
\affiliation{Georgia Institute of Technology}
\author{A. Arulanandan}
\affiliation{Georgia Institute of Technology}
\author{J. Clark}
\affiliation{LIGO Laboratory, Caltech}
\affiliation{Georgia Institute of Technology}
\author{J. Lange}
\affiliation{University of Texas at Austin}
\affiliation{Rochester Institute of Technology}
\author{R. O'Shaughnessy}
\affiliation{Rochester Institute of Technology}
\author{L. Cadonati}
\affiliation{Georgia Institute of Technology}
\begin{abstract}
Gravitational wave parameter inference pipelines operate on data containing unknown sources on distributed hardware with
unreliable performance.   For one specific analysis pipeline (RIFT), we have developed a flexible tool (RUNMON-RIFT)  to mitigate the most
common challenges introduced by these two uncertainties.
On the one hand, RUNMON provides several mechanisms to identify and redress  unreliable computing environments.
On the other hand, RUNMON provides mechanisms to adjust pipeline-specific
run settings, including prior ranges, to ensure the analysis completes and encompasses the physical source parameters.
We demonstrate both general features with two controlled examples.
\end{abstract}
\maketitle

\section{Introduction}

Since the first gravitational wave detection GW150914 \cite{DiscoveryPaper}, the Advanced Laser Interferometer Gravitational Wave Observatory (LIGO)
~\cite{2015CQGra..32g4001L} and  Virgo
\cite{gw-detectors-Virgo-original-preferred,TheVirgo:2014hva}
detectors have
continued to discover gravitational waves (GW) from coalescing binary black holes (BBHs) and neutron stars
\cite{LIGO-O2-Catalog,LIGO-O3-O3a-catalog,LIGO-O3-O3a_final-catalog,LIGO-GW170817-SourceProperties,LIGO-O3-GW190521-implications,LIGO-O3-GW190412,LIGO-O3-NSBH}.
From the point at which data is collected, many computational analyses are required to render it into information of astrophysical interest, including detector characterization \cite{Davis_2021}, calibration and data cleaning\cite{sun2021characterization,virgocollaboration2021calibration}, candidate identification \cite{2016CQGra..33u5004U, CANNON2021100680, mbta}, noise estimation \cite{PhysRevD.91.084034}, and parameter estimation \cite{gwastro-PENR-RIFT-GPU, PhysRevD.91.042003,Ashton_2019}.
For the small number of observations reported through GWTC-2 ($\mathcal{O}(10)$), these analyses could be monitored by individual humans,
to identify and remedy any problems that can occur. 
However, as detector sensitivity improves, the  number of observations and thus inferences increases (potentially to
$\mathcal{O}(100)$ in O4 \cite{LIGO-2013-WhitePaper-CoordinatedEMObserving}), saturating the ability of
individual humans to carefully curate each analysis individually, such that automation will be necessary to correct common problems in future observing runs. This problem is especially acute for parameter inference, which this paper will focus on, though automation schemes have also been implemented for other types of analysis, see for example \cite{Brown_2021,vahi2019custom,gwcelery}, and other work is in progress \cite{daniel_williams_2021_5142601} which deals with similar issues, though with a different focus.

Parameter inference for gravitational waves is generally done within the Bayesian analysis framework. For many possible configurations of parameters which contribute to the gravitational wave (see Section \ref{subsec:RIFT} for details) likelihood values are computed - in this case using approximate models of waveform behavior - and are combined algorithmically with prior expectations to generate posterior distributions which describe the probability of various configurations. For this analysis to be robust, it generally requires at least $\mathcal{O}(10^6)$ likelihood evaluations, which may be computationally expensive. Various methods exist to sample these distributions efficiently, but all are of substantial complexity, and are run primarily on supercomputing clusters. In turn, this complexity allows for many potential issues, both in the settings of the algorithm and the operation of the software, which may drastically reduce the pace of analysis.

 In this paper, we discuss a newly developed Python package, \RM
\footnote{Available at \url{https://git.ligo.org/richard.udall/runmonitor_rift/}}, which seeks to address a number of such problems in inferences performed
using RIFT \cite{gwastro-PENR-RIFT}, one of the parameter estimation (PE) pipelines to interpret events in
GWTC-1 \cite{LIGO-O2-Catalog}, GWTC-2
\cite{LIGO-O3-O3a-catalog}, and GWTC-2.1 \cite{LIGO-O3-O3a_final-catalog}  as well as many individual events
\cite{LIGO-GW170817-SourceProperties,LIGO-O3-GW190521-implications,LIGO-O3-GW190412,LIGO-O3-NSBH}. \RM{} (and \RIFT{} more generally) is geared primarily towards use on the LIGO Data Grid, a collection of independently operated computing clusters running HTCondor \cite{condor-practice,1742-6596-664-6-062014,BOCKELMAN2020101213} with a common software environment and identity access management system, though \RM{} also sees some use on the Open Science Grid \cite{osg07,osg09} via the LDG interface to it.

The challenges faced by software in scientific computing are highly context dependent, and parameter estimation software is no exception. However, some issues are common in many gravitational wave inference pipelines, and we shall focus on discussing these, with solutions tailored to the specific circumstances of RIFT.

A ubiquitous problem with large-scale parameter inference is the presence of computing issues that, though minor and relatively rare, occur
frequently enough in large-scale computation to introduce significant obstacles to automated operation. We introduce tools for managing such issues, both by immediately continuing the progress of a job, and by providing infrastructure to proactively avoid them and provide information about their origin to cluster administrators.

Another common issue with parameter inference is ``railing'': an artificially narrow prior  range that constrains the extent of the posterior distribution in a physical parameter, significantly skewing the final result.
For many practical reasons, PE inference pipelines adopt narrow prior ranges based on expectations informed both by experience and by any additional information, such as the output of a detection pipeline which identified the event originally.
This is imperfect, however, especially when analysis is being done in bulk and available person hours to identify optimal settings are limited.
%
We implement a mechanism for correcting this automatically; this algorithm works best with RIFT for reasons which will be described in more detail in Section \ref{sec:methods}, but could be broadly adapted for any PE software.


This paper is organized as follows.
In Section \ref{sec:methods} we review the RIFT parameter inference engine.
We begin with the core functionalities of \RM{}, including its logging and tools it implements which dramatically decrease the amount of person-hours required to ensure a workflow's completion.
This includes a discussion of common error modes, and of a prototypical computing issue which \RM{} helped overcome.
We then describe how we identify potential 'railing' in our posterior distribution, associated with artificially narrow
boundaries, and we introduce an adaptive method to extend these parameter-space boundaries.
Finally, we discuss a toy model to demonstrate how \RM{} can go beyond reactive workflow management, and proactively ensure that the computational pool used by a workflow is less likely to contain transient computing issues.
Section \ref{sec:results} demonstrates the automated healing of these runs in both a stereotypical case of railing and for our
computing issues toy model.


\section{Methods}
\label{sec:methods}

\subsection{RIFT Review}
\label{subsec:RIFT}


A coalescing compact binary in a quasicircular orbit can be completely characterized by its intrinsic
and extrinsic parameters.  By intrinsic parameters we refer to the binary's  masses $m_i$, spins, and any quantities
characterizing matter in the system.  For simplicity and reduced computational overhead, in this work we provide
examples of parameter inference which assume all
compact object spins are aligned with the orbital angular momentum; however,  the techniques
introduced in our study are not specific to any specific set of parameters or dimension.
By extrinsic parameters we refer to the seven numbers needed to characterize its spacetime location and orientation.
We will express masses in solar mass units and
 dimensionless nonprecessing spins in terms of cartesian components aligned with the orbital angular momentum
 $\chi_{i,z}$.   We will use $\bm{\lambda},\bm{\theta}$ to
refer to intrinsic and extrinsic parameters, respectively.

RIFT \cite{gwastro-PENR-RIFT}
consists of a two-stage iterative process to interpret gravitational wave data $d$ via comparison to
predicted gravitational wave signals $h(\bm{\lambda}, \bm{\theta})$.   In the first stage, denoted by ILE (Integrate Likelihood
over Extrinsic parameters), for each  $\lambda_\alpha$ from some proposed
``grid'' $\alpha=1,2,\ldots N$ of candidate parameters, RIFT computes a marginal likelihood
\begin{equation}
\label{eq:like}\textit{}
 {\cal L}_{\rm marg}\equiv\int  {\cal L}(\bm{\lambda} ,\bm{\theta} )p(\bm{\theta} )d\bm{\theta}
\end{equation}
from the likelihood ${\cal L}(\bm{\lambda} ,\bm{\theta} )$  of the gravitational wave signal in the multi-detector network,
accounting for detector response, and extrinsic parameters prior $p(\bm{\theta})$; see the RIFT paper for a more detailed specification.
In the second stage,  denoted by CIP (Construct Intrinsic Posterior), RIFT performs two tasks.  First, it generates an approximation to ${\cal L}(\bm{\lambda})$ based on its
accumulated archived knowledge of marginal likelihood evaluations
$(\lambda_\alpha,{\cal L}_\alpha)$.  This approximation can be generated by gaussian processes, random forests, or other
suitable approximation techniques.   Second, using this approximation, it generates the (detector-frame) posterior distribution
\begin{equation}
\label{eq:post}
p_{\rm post}=\frac{{\cal L}_{\rm marg}(\bm{\lambda} )p(\bm{\lambda})}{\int d\bm{\lambda} {\cal L}_{\rm marg}(\bm{\lambda} ) p(\bm{\lambda} )}.
\end{equation}
where prior $p(\bm{\lambda})$ is the prior on intrinsic parameters like mass and spin.    The posterior is produced by
performing a Monte Carlo integral:  the evaluation points and weights in that integral are weighted posterior samples,
which are fairly resampled to generate conventional independent, identically-distributed ``posterior samples.''
For further details on RIFT's technical underpinnings and performance,   see
\cite{gwastro-PENR-RIFT,gwastro-PENR-RIFT-GPU,gwastro-mergers-nr-LangePhD}. For managing complex workflows, RIFT
utilizes
  \cite{condor-practice,1742-6596-664-6-062014, BOCKELMAN2020101213} running on a
computing cluster, or the Open Science Grid.

Any  parameter inference analysis generally has many settings, notably including prior assumptions and the amount of
data analyzed. Most relevant to this work is the fact that, for computational efficiency, the priors adopted are generally targeted to cover a limited range of mass and luminosity
distance most likely to enclose the true source parameters, with initial ranges chosen motivated by  search results.
A second critical setting is the starting frequency of the waveform's dominant quadrupole mode.  For an inspiralling
binary at early times, this frequency is twice the orbital frequency.  Because the orbital frequency at the last stable orbit decreases
with mass, for  binaries with a large detector-frame mass,  a conventional starting frequency like $f_{lower} = 20\unit{Hz}$ is too
high: the waveform model doesn't permit it. Furthermore, generation of waveforms with higher modes must start at a
reduced frequency ($f_{min} = 2 f_{lower}/L_{max}$), in order that no mode's initial frequency is above the fiducial starting frequency (i.e., no
mode starts in band).  A third critical setting is the amount of data to analyze, or ``segment
length''.  As the relevant starting frequency or mass decreases, the amount of  data needed to be analyzed increases.
Misidentification of any of these settings can cause a cascade of changes.  For example, a
mis-adapted mass prior might artifically exclude low masses, requiring a re-evaluation of the relevant segment length.


\subsection{\RM{} Introduction}
\RM{} is a Python package to monitor and manage runs,  including
correcting common failure modes encountered.
At its core, \RM{} consists tools to assemble and manage a lightweight run tracking log.
In addition to monitoring the queuing system (condor) logs, \RM{} includes generic tools to parse,  query, and even edit RIFT's internal configuration files and logs.
A daemon will periodically use these tools to update status on each job under its purview, and, using the archived run
logs, the \RM{} suite can quickly assemble reports on run status, including measures of
convergence.
Moreover, being aware of the workflow's status and being able to edit the workflow and even RIFT settings, \RM{} can adapt to issues arising with the host cluster, or individual machines upon it\footnote{A given cluster may have many different machines, with varying behaviors due to hardware architecture, utilization protocols, and the like. We adopt the terminology ``machine" to reflect HTCondor's internal attribute designation, but they may also be variously known as nodes or sites - machines on the primary LDG cluster, for example, have the naming convention node\#\#\#.cluster.ldas.cit.}, in a
fashion that's minimally disruptive to ongoing analysis.
\RM{}'s ``healing'' functions provide unique capability to handle ubiquitous  challenges arising in large-scale
parameter inference calculations.
In this work, we will illustrate  three such operations.   First, we will consider healing parameter ``railing,'' a generic issue
associated with user mis-specified priors.  Second, we will demonstrate how \RM{} can respond to a transient cluster
issue, here exemplified by problems with GPU use.  Finally, we will show how \RM{} can efficiently blacklist undesired
computing machines (e.g., identified by job failures or even slow past performance).


\subsection{Managing Jobs}

In its simplest manifestation, \RM{} implements a run index with operational metadata. LDG clusters feature a web-facing directory which may be
accessed from a browser, and in this directory a file structure is generated, in which the user's run are organized by
what event they are running on. For each run, there are a series of text files containing information about the run such
as the name, location, number of completed iterations, and convergence details, and \RM{} includes a set of utilities which allow the user to parse these conveniently. A daemon is used to automatically analyze the workflows which are registered to its database, and updates the aforementioned metadata accordingly. Thus, we have operational information on all ongoing runs, allowing us to quickly identify potential
problems and characterize overall progress, both critically important when working with many often heterogeneous analyses simultaneously.

\RM{} can provide fixes for some of the many other issues  which can  prevent progress on a run.
These issues can be conveniently flagged by the code, by the use of specific return values from the two key routines
(ILE and CIP).   Alternatively, \RM{} can parse the codes' output and
condor logs, to identify and characterize issues that can cause the run to fail. Quite frequently, these issues are transient in the sense that they are not caused by the structure of the analysis itself, but rather by incompatibilities which occur only in certain parts of a heterogeneous computing pool.

A prototypical example is GPU utilization: \RIFT{} uses GPUs to improve efficiency, but a given cluster may include many
separate machines, often varying dramatically in age. Updates in some standard computing environments to the software
library used by \RIFT{} (\texttt{CUDA}\cite{10.5555/2430671}, called via the Python library
\texttt{CUPY}\cite{cupy_learningsys2017}) rendered it incompatible with some machines on a popular cluster, which in
turn led to an extremely high failure rate, forcing the user to resubmit repeatedly until a job would be lucky enough to
land on a compatible machine. This example motivated the introduction of automated resubmission within \RM{}, so that up
time for runs could be maintained with minimal user intervention, and during times when users would not be
available. Furthermore, it inspired the machine exclusion algorithm described in Section
\ref{subsec:exclusion_description}.
Ultimately, the root of the problem was identified after a number of months, and usage of software libraries was altered
to remove the issue at its source for runs on shared IGWN filesystems, but the resubmission mechanics remain necessary
for the highly heterogeneous OSG pool.  Table \ref{tab:errors} displays several additional errors which RUNMON solves in
an analogous manner.  Many result from instabilities in cluster filesystems which change frequently and are unavoidable for the end user. When a transient is sufficiently common and results in a consistent error message in the Python runtime, \RIFT{} is edited to provide standard error codes for these errors, such that \RM{} may more easily identify and cope with them.

\begin{figure*}[!t]
	\caption{\label{fig:demo:railing}Analysis for the GW190602\_175927 \cite{RICHABBOTT2021100658}, an event where standard parsing of internal low-latency estimates results in initially incorrect boundaries in $\mathcal{M}_c$. Contours are shown for iterations which triggered \RM{}'s railing test, as well as the final result, and vertical lines show the boundaries at the iterations where railing was found. The final boundary occurs substantially to the right of the plot's extent in $\mathcal{M}_c$. Colored points are individual points on the grid, with the heat map corresponding to likelihood.} 
	\includegraphics[width=1.0\linewidth]{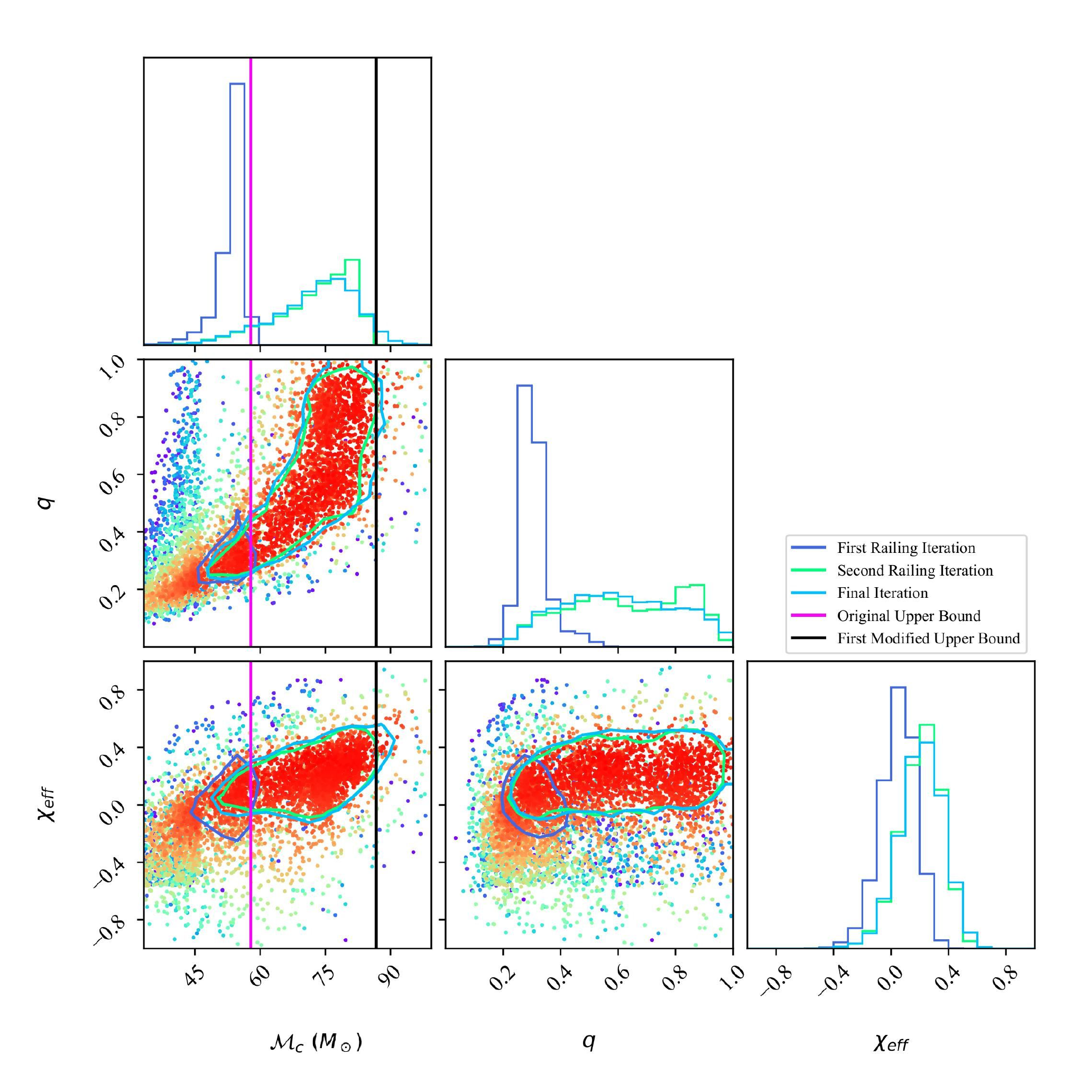}
\end{figure*}

\onecolumngrid
\begin{table}[!h]
\begin{tabular}{c|c|c|c}
	Error Description & Recognition Method & Machine Excludable? & Fixed at Origin? \\ \hline
	CUDA Compute Incompatibility & Custom Error Code & Yes & Yes for IGWN Clusters \\ \hline
	Interpreter Runtime Error & Standard HTCondor Error Code & No & No \\ \hline
	Interpreter Not Found Error & Standard HTCondor Error Code & Yes & Yes \\ \hline
	XLAL File Transient & Output Parsing & No & Yes \\ \hline
\end{tabular}
\caption{\label{tab:errors}Examples of common errors}
\end{table}
\twocolumngrid

\subsection{Healing Parameter Railing}
The priors $p(\bm{\theta}),p(\bm{\lambda})$ over extrinsic and intrinsic parameters are usually proportional to some a priori
seperable function.   In each variable, the range and normalization of the prior is over some finite range.  Sometimes
the boundaries are physical and absolute, for example when integrating over phase or sky location.   However, for
variables like luminosity distance or mass, the user usually adopts upper and lower bounds for computational
convenience, to bound the overall time to solution, centered on a weakly-informed guess.
When performing large-scale inference, these arbitrary bounds are not-infrequently mis-specified, and the posterior is
artificially constrained, ``railing'' against one or more boundaries.


Railing can be identified by having significant posterior support immediately adjacent to one of the arbitrary prior
bounds.
The blue curve in Figure \ref{fig:demo:railing} shows an example of a railed posterior.
To identify railing quantitatively, for some parameter $Z$ we first define a boundary width parameter $B = c(Z_{max}-Z_{min})$, where $c$ is a tunable parameter that we set by default to be $0.05$. Then we consider the points within this boundary width of one of the boundaries:
\begin{equation}
	X_{-} = \{x : Z(x) \in[R_{lower},R_{lower}+B]\}
\end{equation}
\begin{equation}
	X_{+} = \{x : Z(x) \in[R_{upper}-B,R_{upper}]\}
\end{equation}
Then defining $|X_k|$ as the respective cardinality of these sets, we define $P_{\pm} = |X_{\pm}|/N$,
where $N$ is the total number of points in the posterior. If $P > t$, where $t$ is some tunable parameter, defaulting to
$0.03$, then the run is considered to be railed.  [Equivalently, in terms of the default parameters, we say a run is railed if more
  than 3\% of the posterior probability is within the top 5\% of the prior range.]
We emphasize this definition applies only to parameters with \emph{user-specified} boundaries; parameters which have
absolute limits, like the mass ratio $q=m_2/m_1$, do not rail against those limits, as more extreme values are unphysical.


Other parameter estimation methods - notably MCMC and nested sampling methods - require more complex methods of intervention and continuation to achieve similar results, though it also potentially feasible to automate these.  By contrast,  RIFT's intrinsic boundaries  only impact the  second stage calculation [Eq. (\ref{eq:post})],
not the likelihood values portion [Eq. (\ref{eq:like})].
Accordingly, these boundaries may be modified mid-workflow without compromising the analysis. Further iterations are
necessary to populate the grid in the new region, but since railing in masses is normally apparent early in the progress
of an analysis job, the job will in most cases be able to compensate as long as the boundaries are promptly corrected.

Each iteration of RIFT produces a checkpoint posterior, reflecting the distribution from which the next iteration will be sampled. \RM{}'s daemon reads these posteriors, and applies the above algorithm to determine if there is railing in one or both boundary regions. If it is detected, the job is removed from the cluster, and \RM{} changes the boundaries which are found to be railing. If lower bound railing is detected, $R_{lower}$ is mapped to $(1-m)R_{lower}$ and if upper bound railing is detected $R_{upper}$ is mapped to $(1+m)R_{upper}$, where $m$ is another tunable parameter, equal to $0.5$ by default. Once the modification is made, the job is resubmitted and allowed to continue.

\begin{figure*}[!t]
	\caption{The behavior of ILE jobs and number of machines blocked as a function of the associated ILE submission
          batch, for high and low error rate scenarios.  \emph{Left panels}: Histogram of the number of failed jobs
          versus submission attempt. Jobs labelled Succeeded complete normally; jobs noted as RunCrashed have intentionally failed, due
          to landing on a set of pre-selected target hosts; and jobs labelled Crashed fail for other reasons, not
          infrequently associated with problematic or misconfigured host machines.
  \emph{Right panel}: Cumulative number of  blocked as a function of rescue attempt.}
	\includegraphics[width=1\linewidth]{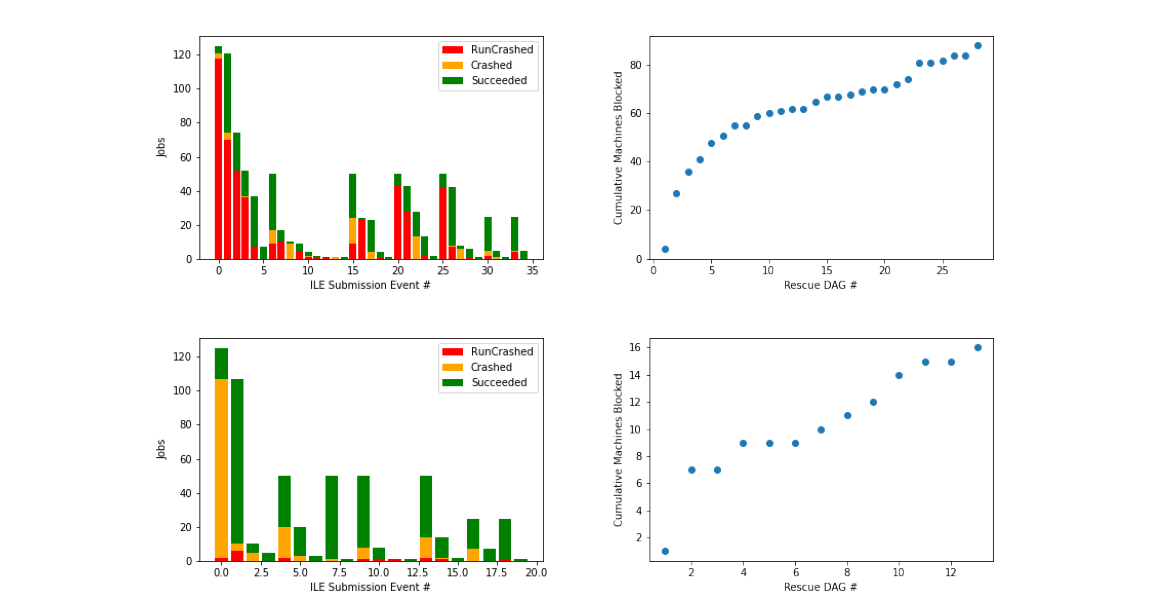}
\end{figure*}

\subsection{Exclusion of Problematic Machines}
\label{subsec:exclusion_description}
Computing clusters frequently suffer from transient errors, usually triggered by some change in the computing
environment, which take the form of everything from failed software dependencies to difficulties with file
transfers. Since the specific conditions required to trigger these transients may only occur for certain computational
tasks, on certain machines, or with specific settings, they may be difficult to track and address. Also problematic is
the phenomena of ``black hole" machines: a colloquial term referring to when a machine will accept a job, but that job will quickly fail due to something inherent to the machine (such as the aforementioned gpu incompatibility). Specifically, if large numbers of jobs are submitted in
parallel (as is the case for high throughput computing (HTC) tasks, such as the ILE stage of RIFT), the scheduler will
attempt to assign them in bulk. If the available computational resources are limited, then some fraction will be
assigned and the rest will occupy the next spots in the queue. Accordingly, a single machine experiencing some transient
error may fail immediately, then accept another job from this queue. In sufficiently low resource situations, this may
result in the entire parallel content of an HTC job failing on a single machine.

To mitigate the impact of problematic machines, \RM{} allows for
tracking machines associated with known transient errors, and provides a tool for instructing the relevant condor jobs
to exclude these machines from matchmaking consideration. The information about which machines should be excluded is
shared across all jobs managed by a given daemon, and thus is propagated quickly for all of a user's active
jobs. Logging of these machines and their associated failure modes also offers a collated set of data to provide system
administrators when troubleshooting issues, such that the root problem may be identified and addressed, at which point
it is straightforward to remove the restrictions the daemon imposed upon the pool.


\section{Results}
\label{sec:results}

\subsection{Healing}

Figure \ref{fig:demo:railing} depicts a prototypical example of railing, along with the correction produced by \RM{}. The pipeline constructor for the event in question, GW190602\cite{RICHABBOTT2021100658}, produced a railed prior boundary in chirp mass $\mathcal{M}_c$ when taking the metadata of the event's initial detection as input. Accordingly, it required careful and tedious human intervention, lest any run be completely ruined. The use of \RM{} may be seen to alleviate this in the progression of the results seen in Figure \ref{fig:demo:railing}. The plot in question is an example of a corner plot - displaying both one dimensional histograms of individual parameters, as well as their two dimensional joint parameters, so that correlation may be understood and diagnosed. \RIFT{} corner plots also include colored points to show the likelihood values of the underlying grid, with the hot colors corresponding to the highest likelihood points, and the cool colors corresponding to the lowest likelihood points. Here the posterior after the first iteration of the workflow (the blue curve) may be seen to rail at the upper boundary in $\mathcal{M}_c$ which was set by the pipeline constructor (the pink line). Notably, this distribution also has an erroneous posterior distribution in mass ratio $q$, due to the correlation of this parameter with the erroneous chirp mass. \RM{} then automatically increased the upper boundary to the value seen in the black line, and continued the sampling process. After a number of iterations, the posterior had shifted to the seafoam curve, which may be seen to also rail (though to a lesser degree) against the modified upper bound, and so \RM{} modified the upper bound once more, well past limits of the corner plot ($\mathcal{M}_c \approx 135 M_\odot$). Final sampling then brought the posterior to an unrailed distribution (the teal curve), which agrees with the results presented for this event in \cite{LIGO-O3-O3a-catalog}.



\subsection{Runcrasher}
To demonstrate the principle of machine exclusion, we construct an artificial scenario with known parameters and behavior which mimics the transient errors known to occur on computational clusters. In particular, we insert a step into the standard ILE portion of the workflow which tests the machine upon which the job lands, and produces a failure if that machine's name satisfies certain constraints (e.g. if the last digit is 5). \RM{} included this failure code as one of the known transient values, and the exclusion system was triggered accordingly. We conducted tests under various constraints, reflecting the varying incidence rate of transients. This construct also naturally results in the aforementioned ``black hole" machines when submission incidentally occurs at a time of high resource usage.

The results of this artificial scenario and corresponding intervention are shown in Figure 2. The bar charts indicate the behavior of individual machines under high and low transient incidence rates respectively. Transients are separated into two types: those which are caused by the runcrasher, which behave in a predictable manner and are subject to machine exclusion, and those which are caused by miscellaneous other transients, which are not well characterized and not subject to machine exclusion (for the runs in question these transients primarily involve accessing certain public files). The scatter plots show how many machines are actively excluded for each of these corresponding submission batches.

A number of features may be noted in these plots. Firstly, the submission events for which the total number of jobs
increase are those submissions which occur at the beginning of a new iteration.  The total duration of iterations for
which the same numbers of jobs are submitted decrease correspondingly in the high-incidence case (the number of jobs
submitted per iteration varies over the duration of the underlying workflow to improve its efficiency). Similarly, the relative proportion of errors which are due to the unmodeled transients increases. The low
incidence case shows somewhat similar behavior, though it is also more strongly subject to low number statistics, as it
is relatively rare to hit a failure machine in the first place.

When including analysis of the number of machines submitted, one may also see the expected trend: initial jobs result in substantially more blocked machines, while later jobs run in a cleaner pool, and thus are less likely to simultaneously interact with many error-triggering machines. One may also note that there are submission events (submission event 20, for example) for which most jobs fail but very few machines are blocked - this is an example of the aforementioned black-hole machine phenomenon. These unfortunately take longer to root out, since it gets progressively harder to filter through the pool when it is already mostly successful, but integration of these error lists across multiple runs mean that in a high usage context (if one has 10 runs simultaneously, for example) it is very feasible to fully eliminate problematic machines.




\section{Conclusion}

We have presented our Runmonitor for RIFT (\RM{}), a utility which greatly aids in the operation of this inference pipeline. Gravitational wave science is in an exciting time, with a rapid pace of discovery and exponentially increasing data to analyze. In this context, it is critical that the time required to complete a parameter estimation task, and the time the user spends actively monitoring and intervening in that task, be minimized. By introducing centralized diagnostic tools, \RM{} makes it much easier for a user to check the status of their active jobs. Automated resubmission for known transient issues greatly decreases the amount of time a user spends actively engaging with the workflow (in particularly hostile computing environments this decrease may be up to an order of magnitude), and machine exclusion allows one to tailor the pool utilized towards the machines which will actually work consistently, decreasing restarts and improving efficiency for all cluster users. Monitoring of railing allows for aggressive (and hence efficient) initial settings, while also reducing the need for producing new workflows during exploratory phases of analysis.

\section{Acknowledgments}

The authors are grateful for computational resources provided by the LIGO-Laboratory, Leonard E Parker Center for Gravitation, Cosmology and Astrophysics at the University of Wisconsin-Milwaukee, and Inter-University Center for Astronomy \& Astrophysics (IUCAA), Pune, India, and supported by National Science Foundation Grants PHY-1626190, PHY-1700765, HY-0757058 and PHY-0823459.
This research was done using resources provided by the Open Science Grid \cite{osg07,osg09}, which is supported by the National Science Foundation award \#2030508.
This research has made use of data, software and/or web tools obtained from the Gravitational Wave Open Science Center (https://www.gw-openscience.org/ ), a service of LIGO Laboratory, the LIGO Scientific Collaboration and the Virgo Collaboration. LIGO Laboratory and Advanced LIGO are funded by the United States National Science Foundation (NSF) as well as the Science and Technology Facilities Council (STFC) of the United Kingdom, the Max-Planck-Society (MPS), and the State of Niedersachsen/Germany for support of the construction of Advanced LIGO and construction and operation of the GEO600 detector. Additional support for Advanced LIGO was provided by the Australian Research Council. Virgo is funded, through the European Gravitational Observatory (EGO), by the French Centre National de Recherche Scientifique (CNRS), the Italian Istituto Nazionale di Fisica Nucleare (INFN) and the Dutch Nikhef, with contributions by institutions from Belgium, Germany, Greece, Hungary, Ireland, Japan, Monaco, Poland, Portugal, Spain."
This material is based upon work supported by NSF’s LIGO Laboratory which is a major facility fully funded by the National Science Foundation.

\bibliography{references,LIGO-publications}
\end{document}